\input ./style/arxiv-general.cfg
\documentclass[aoas,MSNbibl,nameyear,dvips]{arximspdf}
\makeatletter
   \@ifpackageloaded{graphicx}{}{\usepackage{graphicx}}
\makeatother

\doi{10.1214/15-AOAS840}
\volume{9}
\issue{3}
\pubyear{2015}
\firstpage{1549}
\lastpage{1570}
\docsubty{FLA}


\begin{document}
\begin{frontmatter}

\title{Combining nonexchangeable functional or survival data sources in
oncology using generalized mixture commensurate priors}
\runtitle{GMC prior models}

\begin{aug}
\author[A]{\fnms{Thomas A.}~\snm{Murray}\thanksref{M1,T1}\ead[label=e1]{tamurray@mdanderson.org}},
\author[A]{\fnms{Brian~P.}~\snm{Hobbs}\thanksref{M1,T1,T2}}~\and\\
\author[B]{\fnms{Bradley~P.}~\snm{Carlin}\corref{}\thanksref{M2,T1}\ead[label=e3]{carli002@umn.edu}}
\runauthor{T. A. Murray, B. P. Hobbs and B. P. Carlin}
\affiliation{University of Texas MD Anderson Cancer
Center\thanksmark{M1} and\break University of Minnesota\thanksmark{M2}}
\address[A]{T. A. Murray\\
B. P. Hobbs\\
Department of Biostatistics---Unit 1411 \\
The University of Texas\\
\quad MD Anderson Cancer Center\\
P.O. Box 301402 \\
Houston, Texas 77230\\
USA\\
\printead{e1}}
\address[B]{B. P. Carlin\\
Division of Biostatistics \\
School of Public Health\\
 University of Minnesota \\
420 Delaware St SE MMC 303 \\
Minneapolis, Minnesota 55455\\
USA\\
\printead{e3}}
\end{aug}
\thankstext{T1}{Supported in part by NCI Grant 1-R01-CA157458-01A1.}
\thankstext{T2}{Supported in part by Cancer Center Support Grant (CCSG)
(P30 CA016672).}

%
\received{\smonth{3} \syear{2014}}
%
\revised{\smonth{3} \syear{2015}}

\begin{abstract}
Conventional approaches to statistical inference preclude structures
that facilitate incorporation of supplemental information acquired from
similar circumstances. For example, the analysis of data obtained using
perfusion computed tomography to characterize functional imaging
biomarkers in cancerous regions of the liver can benefit from partially
informative data collected concurrently in noncancerous regions. This
paper presents a hierarchical model structure that leverages all
available information about a curve, using penalized splines, while
accommodating important between-source features. Our proposed methods
flexibly borrow strength from the supplemental data to a degree that
reflects the commensurability of the supplemental curve with the
primary curve. We investigate our method's properties for nonparametric
regression via simulation, and apply it to a set of liver cancer data.
We also apply our method for a semiparametric hazard model to data from
a clinical trial that compares time to disease progression for three
colorectal cancer treatments, while supplementing inference with
information from a previous trial that tested the current standard of care.
\end{abstract}

\begin{keyword}
\kwd{Bayesian hierarchical model}
\kwd{clinical trials}
\kwd{colorectal cancer}
\kwd{commensurate prior}
\kwd{computed tomographic imaging}
\kwd{evidence synthesis}
\kwd{mixture priors}
\kwd{penalized splines}
\kwd{proportional hazards}
\kwd{semiparametric methods}
\end{keyword}
\end{frontmatter}

\section{Introduction}\label{SIntro}

Statistical investigations begin by determining which  sources(s) of
information will be used to answer the motivating questions and
generate hypotheses for future exploration. Conventional approaches to
statistical inference preclude structures that facilitate incorporation
of partially informative data, imposing polarity on the data selection
process. Putatively relevant supplemental data acquired from broadly
similar therapeutic interventions, patient cohorts, previous
investigations or biological processes are often either excluded from
statistical analysis or treated as exchangeable with the primary data.
In oncology, prospective studies designed to evaluate the performance
of an experimental therapy usually ignore historical information about
the control therapy and limit enrollment to patients presenting lesions
with a particular histological subtype, grade or performance status, or
who are na\"{i}ve to prior therapy. In contrast, studies with more
liberal information inclusion rules, such as intention-to-treat
designs, often pool information from potentially heterogeneous sources
or adjust for this potential heterogeneity using simple linear regressors.

Ignoring relevant, supplemental sources of information reduces the
reproducibility and scope of the primary study. On the other hand,
using the supplemental information while neglecting to account for
heterogeneity between the sources of information obscures understanding
of the complex underlying mechanisms that produced the primary data and
may lead to severely biased inference. Relaxing this dichotomy would
improve the efficiency of the experimental process and enable
investigators to implement statistical models that use all available
information while accommodating important between-source features.
Several models have been proposed for incorporating partially
informative supplemental data. \citet{poco76} used generalized linear
models with static, data-independent borrowing using a prespecified
amount of between-source variability. \citet{ibra00} proposed
data-independent or dynamic nonhierarchical methods for partially
weighting likelihoods. Bayesian [\cite{smit95}] and frequentist
[\cite{doi11}] methods using hierarchical modeling have been developed for
estimating between-source variability for univariate observables or
repeated measures with generalized linear relations among covariates.

Bayesian hierarchical models facilitate dynamic partial pooling of
between-source information and flexibly estimate the extent of
borrowing. Here the goal is limiting bias for estimating primary
effects when between-source heterogeneity occurs, while improving
efficiency when approximate coherence, or \textit{commensurability},
occurs; see \citet{hobb12} for generalized linear mixed models, and
\citet{hobb13} and \citet{murr14} for piecewise-exponential
time-to-event models. In Section~\ref{SSupp} we discuss these methods,
which can be used to borrow strength from supplemental data in
parametric models for regression coefficients and other univariate
parameters. We then develop a Bayesian hierarchical model structure to
leverage supplemental information more generally, including for both
nonparametric regression and semiparametric hazard models with
penalized splines. The literature appears devoid of general methods for
flexibly borrowing strength from supplemental data in semi- and
nonparametric models for a group of related parameters that
characterize a complex object (e.g., a curve or surface).

Two oncological applications motivate our methodological developments.
The first, from diagnostic radiology, involves estimating prognostic
functional imaging biomarkers acquired using perfusion computed
tomography (CT) in cancerous and cancer-free liver tissue. Perfusion CT
is an emerging technology that enables observation and quantification
of characteristics pertaining to the passage of fluid through blood
vessels. Researchers have developed physiological models to quantify a
variety of perfusion characteristics derived from analysis of the
distribution of contrast enhancement in tissue acquired using repeated
CT scans during intravenous administration of contrast medium.
Investigators have used the technology in a number of organs and
tumors, including prostate, colorectal, head and neck, lung and liver.
\citet{mile03} review the clinical relevance of perfusion CT.

\begin{figure}[t]

\includegraphics{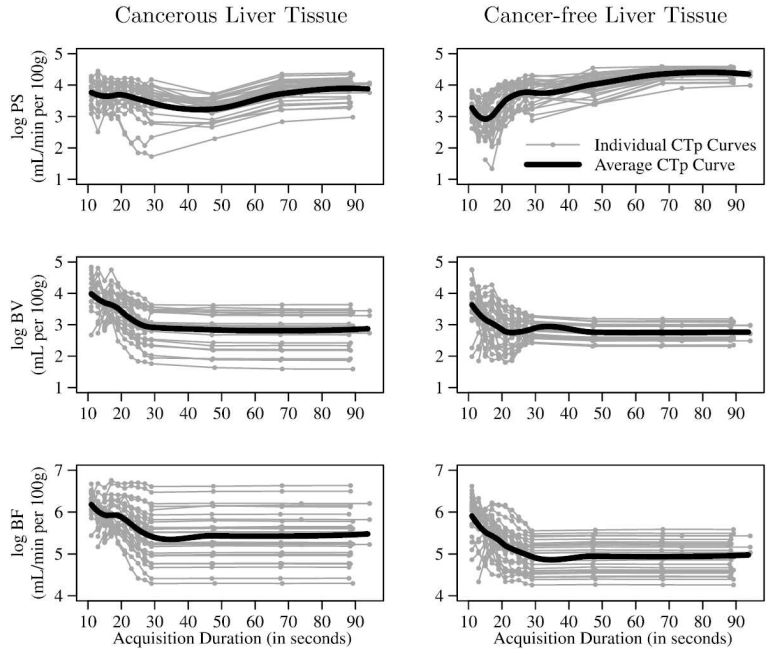}

\caption{Individual-by-region perfusion CT (CTp) curves are displayed
by tissue region (``Individual CTp curves'') for permeability-surface
area product (PS), blood volume (BV) and blood flow (BF), over
acquisition durations between 11 and 95 seconds after contrast
injection. The dots mark actual observations. The thick ``Average CTp
curve'' is a Loess estimate that ignores potential clustering.}\label
{PRawCTpPlot}
\end{figure}

Our application considers three characteristics: permeability-surface
area product (PS), blood volume (BV), and blood flow (BF). Each
characteristic is measured at 7 to 13 acquisition times between 11 and
95 seconds following contrast injection. Data from 16 individuals
comprise a total of 25 regions containing pathology-verified metastases
to the liver from neuroendocrine tumors (i.e., cancerous liver tissue),
and a total of 27 regions consisting of noncancerous liver tissue. PS
and BF are rates measured as milliliters per minute per 100 grams of
liver tissue (ml/min per 100~g), whereas BV is a volume measured as
milliliters per 100 grams of liver tissue (ml per 100~g). Figure~\ref{PRawCTpPlot} displays the observed perfusion CT (CTp) curves along
with Loess estimates of the average CTp curve for each characteristic
in each tissue region. Characterization of the perfusion
characteristics in cancerous tissue has implications for constructing
biomarkers to assist in treatment monitoring, prognostication and
pathophysiological understanding of metastatic tumor vasculature. It is
important that the acquisition duration cover the range of CTp curve
instability, but once the curve has stabilized, the scan can be
terminated. To limit radiation exposure and cost, the acquisition
duration should be minimized [\cite{ng13}]. Critically, because the
tissue type is unknown prior to diagnosis, any proposed acquisition
period must ensure stable quantification of CTp characteristics for
both types of tissue before CTp can be used for detection of metastatic
sites. Though nonexchangeable, the perfusion CT data obtained in
cancer-free liver tissue may inform the shape and stabilization time of
the corresponding CTp curve in cancerous tissue. Figure~\ref{PRawCTpPlot} shows substantial heterogeneity in the shapes of the
average CTp curves by tissue region for PS, whereas for BV and BF, the
average CTp curves are more similar.

Our second application, from colorectal cancer, involves estimating
prog\-ression-free survival (PFS) with data from two consecutive
randomized phase III colorectal cancer trials, reported by \citet
{salt00} and \citet{gold04}. Both trials used PFS to assess the
efficacy of various treatment regimens for patients with previously
untreated metastatic colorectal cancer; disease progression was defined
as a 25\% increase in measurable tumor size, presence of a new lesion
or death. The initial trial [\cite{salt00}] compared three treatment
regimens: 5-Fluorouracil and Leucovorin, Irinotecan alone, and
Irinotecan and bolus Fluorouracil plus Leucovorin (IFL). The results
indicated that the IFL regimen was significantly more efficacious than
the other two regimens, and IFL became the ``standard of care'' leading
into the subsequent trial [\cite{gold04}], which then compared an
identical IFL regimen with two novel regimens: Oxaliplatin with infused
Fluorouracil plus Leucovorin (FOLFOX), and Irinotecan with Oxaliplatin (IROX).

\begin{figure}

\includegraphics{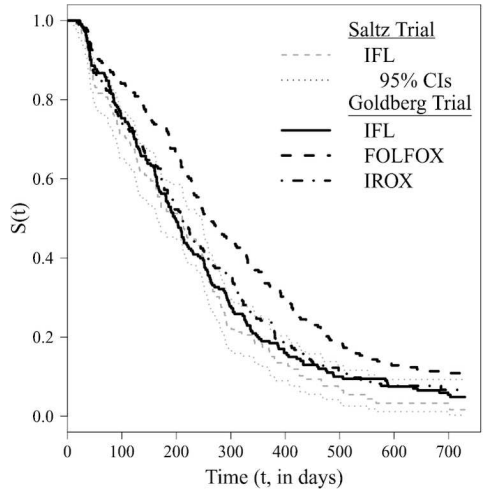}

\caption{Kaplan--Meier estimates of PFS for the IFL regimen in the Saltz
trial (along with 95\% CIs), and the IFL, FOLFOX and IROX regimes in
the Goldberg trial.}\label{PRawKMPlot}
\end{figure}

Figure~\ref{PRawKMPlot} shows that the PFS curves for the IFL regime
are commensurate in the Saltz and Goldberg trials, with the PFS curve
in the Goldberg trial tracking just above that of the Saltz trial,
though within the 95\% CIs for nearly all of follow-up. Moreover,
Figure~\ref{PRawKMPlot} suggests that in the Goldberg trial FOLFOX is
superior to both IROX (log-rank test $p$-value of 0.006) and IFL
(log-rank test $p$-value${}<0.001$), and that IFL and IROX perform
similarly (log-rank test $p$-value of 0.404). The efficiency for
estimating the PFS curve of the IFL regimen in the Goldberg trial may
be improved by using nonexchangeable, yet relevant data on the IFL
regimen in the Saltz trial.

The commonality between our two motivating examples is the availability
of supplemental information about unknown curves that may aid our
inference in the primary investigation (i.e., CTp curves among tissue
types, and PFS curves of IFL among clinical trials). In Section~\ref{SNPReg} we investigate the borrowing properties via simulation of the
proposed hierarchical model structure for nonparametric regression
using penalized splines, and then we analyze the perfusion CT data. In
Section~\ref{SHazReg} we analyze data from the colorectal cancer
clinical trials using the proposed hierarchical structure for a
semiparametric hazard model. In Section~\ref{SDisc} we close with a
discussion and propose directions for future work.

\section{Leveraging supplemental information}\label{SSupp}
We restrict our attention to two Bayesian methods for leveraging
supplemental information, power priors [\cite{ibra00}] and commensurate
priors [\cite{hobb11}]. In general, Bayesian models consist of a
likelihood for the data and prior specifications for the parameters in
the likelihood [see, e.g., \cite{carl09}]. For both these modeling
approaches, the \textit{primary} and \textit{supplemental} likelihoods are
assumed to have the same structure (e.g., both Gaussian), and we denote
them by $L(\bolds{\theta}|\mathbf{D})$ and $L(\bolds{\theta}_0|\mathbf{D}_0)$,
respectively. As a result, the supplemental parameter $\bolds{\theta}_0$
is analogous to the primary parameter $\bolds{\theta}$, and information
about $\bolds{\theta}_0$ can be leveraged to aid posterior inference about
$\bolds{\theta}$.

\subsection{Existing methods}\label{SSCommPrior}
Power priors assume the sources of information are exchangeable (i.e.,
$\bolds{\theta} \equiv\bolds{\theta}_0$) and downweight the supplemental
likelihood by raising it to a prespecified power $a_0 \in[0,1]$. This
strategy works because raising the supplemental likelihood to the power
$a_0$ diffuses it. In the extreme case where $a_0 = 0$, the
downweighted supplemental likelihood is a noninformative constant.
Formally, the posterior arises as
\begin{equation}
\label{EPowPr}
p(\bolds{\theta}|\mathbf{D},\mathbf{D}_0) \propto L(
\bolds{\theta}|\mathbf{D}) L(\bolds{\theta}|\mathbf{D}_0)^{a_0}
\pi(\bolds{\theta}).
\end{equation}
Alternatively, $a_0$ can be modeled as another unknown parameter in the
model [\cite{ibra00}]. Power priors are extremely general, so nothing
explicitly prevents their use in any modeling context. However, they
employ a single parameter ($a_0$) to control between-source borrowing,
thereby making differential borrowing among components of $\bolds{\theta}$
infeasible unless $L(\bolds{\theta}|\mathbf{D})$ has a convenient
factorization, as it generally does not. This is undesirable because
the supplemental data may provide relevant information for only a
subset of $\bolds\theta$. Furthermore, when $a_0$ is assumed unknown,
power priors tend to excessively downweight the supplemental
likelihood, even when $\mathbf{D}_0$ and $\mathbf{D}$ are identical [\cite{neel10}].

Commensurate priors do not assume the sources of information are
exchangeable (i.e., $\bolds{\theta} \not\equiv\bolds{\theta}_0$), and
instead specify a hierarchical model where the posterior arises as
\begin{equation}
\label{ECommMod}
p(\bolds{\theta},\bolds{\theta}_0,\bolds{\eta}|
\mathbf{D},\mathbf{D}_0) \propto L(\bolds{\theta}|\mathbf{D}) L(
\bolds{\theta}_0|\mathbf{D}_0) \pi(\bolds{\theta}|
\bolds{\theta}_0,\bolds{\eta}) \pi(\bolds{\eta}) \pi(\bolds{\theta}_0).
\end{equation}
The \textit{commensurate prior}, $\pi(\bolds{\theta}|\bolds{\theta}_0,\bolds{\eta
})$, ``centers'' $\bolds{\theta}$ about $\bolds{\theta}_0$, and $\bolds{\eta}$
controls between-source borrowing to reflect the commensurability of
the two parameters. In practice, $\pi(\bolds{\theta}|\bolds{\theta}_0,\bolds{\eta
})$ is defined so that $E[\bolds{\theta}|\bolds{\theta}_0,\bolds{\eta}] = \bolds{\theta}_0$ and $\operatorname{Var}[\bolds{\theta}|\bolds{\theta}_0,
\bolds{\eta}]$ is
decreasing in $\bolds{\eta}$.

For a unidimensional real-valued parameter $\theta$, a useful approach
takes $\theta|\theta_0,\eta\sim\mathcal{N}(\theta_0,\eta^{-1})$, a
Gaussian distribution with mean $\theta_0$ and precision $\eta$.
Estimation of $\eta$ is inherently difficult, but feasible by inducing
sparsity over the precision domain using a ``spike-and-slab'' prior for
$\eta$ [\cite{hobb12}]. The spike-and-slab prior has a mixture density
\begin{equation}
\label{ESpSl1}
\pi(\eta) \equiv(1-p_0) \mathcal{U}(
\eta|s_l,s_u) + p_0 \delta _{\mathcal{R}}(
\eta),
\end{equation}
where $0 \le s_l < s_u << \mathcal{R}$ and $p_0 \in[0,1]$ are
prespecified, and $\delta_{\mathcal{R}}(\eta)$ is one at $\mathcal{R}$
and zero otherwise. The distribution in (\ref{ESpSl1}) is locally
uniform on the ``slab,'' $(s_l,s_u)$, with probability $(1-p_0)$, and
places probability mass $p_0$ at the ``spike,'' $\mathcal{R}$, otherwise.

We prefer to modify the commensurate prior construction suggested by
\citet{hobb12} as
\begin{eqnarray}
&&\theta|\theta_{0},\tau,\iota\sim \bigl[\mathcal{N}
\bigl(\theta _{0},\tau^{-1}\bigr) \bigr]^{(1-\iota)}
\bigl[\mathcal{N}\bigl(\theta _{0},\mathcal{R}^{-1}\bigr)
\bigr]^{\iota},
\nonumber
\\[-8pt]
\label{ESpSl2}
\\[-8pt]
\eqntext{\displaystyle\iota\sim \operatorname{Bern}(p_0) \mbox{ and } \tau\sim
\mathcal{U}(s_l,s_u),}
\end{eqnarray}
where $\operatorname{Bern}(p_0)$ denotes a Bernoulli distribution with $\operatorname{Pr}(\iota= 1)
= p_0$. Therefore, the commensurate prior in (\ref{ESpSl1}) induces a
two-part mixture prior distribution on~$\theta$ that consists of a
highly concentrated component [i.e.,\break $\mathcal{N}(\theta_{0},\mathcal
{R}^{-1})$] and a relatively diffuse component [i.e., $\mathcal
{N}(\theta_{0},\tau^{-1})$], both centered at $\theta_0$. The model in~(\ref{ESpSl2}) has parallels to Bayesian variable selection methods,
in which the two-part mixture prior for a regression coefficient has
one component heavily concentrated about zero and the other vague [see,
e.g., \cite{geor97}]. Hence, the commensurate prior facilitates
selective borrowing by inheriting the selective shrinkage property of
the spike-and-slab distributions used in Bayesian variable selection.
In particular, the commensurate prior strongly shrinks $\theta$ to
$\theta_0$ when evidence indicates that this difference is small,
thereby improving efficiency, and minimally affects $\theta$ otherwise,
thereby limiting bias.

We specify the spike so that if $\theta|\theta_0 \sim\mathcal{N}(\theta
_{0},\mathcal{R}^{-1})$, $\theta$ deviates negligibly from~$\theta_0$.
On the other hand, we specify the slab to contain small values that
correspond to modest shrinkage of $\theta$ toward $\theta_{0}$. Hence,
a $\mathcal{N}(\theta_{0},s_u^{-1})$ prior for $\theta$ will be weakly
informative. The tails of the prior distribution in (\ref{ESpSl2}) are
jointly controlled by $s_l$ and $s_u$: when $s_l \approx0$, smaller
values of $s_u$ provide a prior with heavier tails. In our experience,
posterior inference on $\theta$ is insensitive to modest shifts in
these hyperparameters, so we suggest specifying $s_l$ near zero, with
$s_u$ small and $\mathcal{R}$ large relative to the magnitude of a
meaningful difference between $\theta$ and $\theta_0$. Posterior
inference \textit{can} be sensitive to the fourth hyperparameter, $p_0$,
the prior probability of effective equality between $\theta$ and $\theta
_0$. We recommend choosing the value for $p_0$ to deliver satisfactory
operating characteristics at the anticipated sample size, such as
mean-squared-error properties over a range of likely true differences
between $\theta$ and $\theta_0$; see \citet{murr14}.

Specifying a value for $s_u$ that is too large can result in a model
that is not robust to between-source heterogeneity, whereas specifying
a value for $\mathcal{R}$ that is too small will result in a model that
gains little posterior precision for $\theta$ even when $\mathbf{D}_0$ and
$\mathbf{D}$ are equivalent. Specifying too small a value for $p_0$ may
also result in a model that borrows little when $\mathbf{D}_0 \equiv\mathbf{D}$, whereas specifying too large a value for $p_0$ may result in a
model that borrows too much (resulting in unacceptable bias) in the
presence of substantial between-source heterogeneity. Of course,
$p_0$'s influence on posterior inference diminishes as the primary
sample size increases.

\subsection{Generalized mixture commensurate priors}\label{SSGMCPrior}
The general model structure described in (\ref{ECommMod}) can
accommodate a latent precision parameter (component of $\bolds{\eta}$) for
each component of $\bolds\theta$. Therefore, the extent to which the
supplemental data influence estimation of the primary effects can be
allowed to differ among the components of $\bolds\theta$, yielding
flexibility and reducing bias. However, assuming mutual independence
between the components of $\bolds{\eta}$ allows strength to be borrowed
from the supplemental source(s) differentially among the components of
$\bolds\theta$, reducing efficiency. Though this may be sensible,
differential borrowing among the components of $\bolds\theta$ may not
always be appropriate.

Suppose $\bolds{\beta} = (\beta_1,\ldots,\beta_K)$ is a subset of $\bolds
{\theta}$ that characterizes a feature in the primary data (e.g., the
shape of a CTp curve), with an analogous definition for $\bolds{\beta}_0$.
We may wish to borrow from the supplemental data for $\bolds{\beta}_0$ to
a degree that reflects its coherence with $\bolds{\beta}$ in its entirety,
rather than case-by-case for each $\beta_k$. Moreover, $\bolds{\beta}$
(and $\bolds{\beta}_0$) may conventionally be assigned a prior that
induces smoothness. Ideally, (\ref{ECommMod}) will inherit the
smoothness properties of the conventional prior and simultaneously
borrow similar amounts for related subgroups of $\bolds{\theta}$ (i.e.,
$\bolds{\beta}$). We propose to model
\begin{eqnarray}
\beta_k|
\iota_k,\beta_{k,0}  &\sim&  \bigl[\pi^*(\beta_k|
\beta _{k,0}) \bigr]^{1-\iota_k} \bigl[\pi(\beta_k|\beta
_{k,0}) \bigr]^{\iota_k},\nonumber\\[-8pt]
\label{EGenMixCommPr}
\\[-8pt]
 \eqntext{\displaystyle\iota_k|\nu\sim \operatorname{Bern}(\nu),
\nu\sim\mathcal{B}(a_1,a_2), \beta_{0,k} \sim
\pi^*(\beta_{0,k}),}
\end{eqnarray}
where\vspace*{1pt} $\mathcal{B}(a_1,a_2)$ denotes a beta distribution with mean
$\frac{a_1}{a_1+a_2}$ and the specifications for $\pi^*$ mirror a
conventional analysis of these data. To facilitate borrowing of
strength, we take $\pi(\beta_k|\beta_{0,k})$ in (\ref{EGenMixCommPr})
to be heavily concentrated about $\beta_{0,k}$; we provide exact
specifications for each application later. To borrow similar amounts of
strength for each component of $\bolds{\beta}$, we assume the $\iota_k$
are identically distributed with $\operatorname{Pr}(\iota_k=1) = \nu$. Thus, the
resulting prior is composed of $K$ two-piece mixtures and generalizes
the commensurate prior defined in (\ref{ESpSl2}). Hereafter, we refer
to the general prior structure defined in (\ref{EGenMixCommPr}) as a
\textit{generalized mixture commensurate} (GMC) prior.

\section{GMC priors in nonparametric regression analysis}\label{SNPReg}
Suppose $\mathbf{y} = (y_1,\break \ldots,y_N)$ is a real-valued variable (e.g.,
log blood flow, as in Figure~\ref{PRawCTpPlot}) that depends on a
continuous covariate $\mathbf{t} = (t_1,\ldots,t_N)$ (e.g., time, as in
Figure~\ref{PRawCTpPlot}), and that we model $y_i \sim \mathcal
{N}(\phi(t_i),\sigma^2)$. Often primary interest lies in the shape,
derivatives or some other function of $\phi$, and $\phi$ requires a
smooth but flexible nonparametric specification, as opposed to a
parametric linear specification $\phi(t) = \beta_0 + \beta_1 t$.
Penalized splines are a practical choice for modeling $\phi$; see \citet{rupp03}, Chapters~3 and 14.
There are many ways to formulate a
penalized spline. Low-rank thin-plate (LRTP) splines are appealing
since they are defined by tractable radial basis functions and tend to
exhibit fast Markov chain Monte Carlo (MCMC) convergence properties in
a Bayesian context relative to truncated basis splines [\cite{crai05}].
B-splines are another reasonable Bayesian option because they too tend
to exhibit fast MCMC convergence, but they rely on a recursive
algorithm to define the basis functions, making them less tractable
than LRTP splines [\cite{eile96}]. Our method will work with either
LRTP or B-spline formulations for $\phi$, and we do not anticipate
substantial differences in the behavior of our method under either
formulation. Hereafter, we focus on a LRTP cubic spline specification
for $\phi$.

Without loss of generality, we take $t \in[0,1]$ and model
\begin{equation}
\label{EmLRTP}
\phi(t;\bolds{\beta}) = \beta_0 +
\beta_1 t + \sum_{k=2}^{K}
\beta _k \bigl(|t-\tilde{t}_{k-1}|^3 - |
\tilde{t}_{k-1}|^3 \bigr),
\end{equation}
where $\tilde{\mathbf{t}} = (0=\tilde{t}_0<\tilde{t}_1<\cdots<\tilde
{t}_{K-1}<\tilde{t}_K = 1)$ is a generic partition with $K$ intervals.
We discuss this choice later.
Typically, a LRTP spline model is formulated as
\begin{equation}
\label{ELRTP} \phi\bigl(t;\bolds{\beta}^*\bigr) = \beta^*_0 +
\beta^*_1 t + \sum_{k=2}^{K}
\beta^*_{k}|t-\tilde{t}_{k-1}|^3;
\end{equation}
however, in (\ref{ELRTP}), $\beta^*_0$ is not an intercept because
$\phi(0;\bolds{\beta}^*) = \beta^*_0 + \sum_{k=2}^{K} \beta
^*_{k}|\tilde{t}_{k-1}|^3$. We prefer the modified LRTP (mLRTP) model
defined in (\ref{EmLRTP}), because it simplifies interpretation and
prior elicitation when the regression includes an intercept, which will
be the case for the CTp data, and it retains an equivalent
interpretation of $(\beta_1,\ldots,\beta_K)$ and $(\beta^*_1,\ldots
,\beta^*_K)$.
In the sequel, we denote $\bolds{\beta}_{(-0)} = (\beta_1,\ldots,\beta
_K)$, that is, the vector $\bolds{\beta}$ with $\beta_0$ omitted.

A \textit{penalized} spline requires prespecification of a fine partition
$\tilde{\mathbf{t}}$, and smooths via the prior, whereas a \textit{free-knot}
spline uses complex sampling algorithms (e.g., reversible-jump MCMC) to
sample over all possible partitions. We prefer penalized splines, since
they are computationally much simpler and have proven to be competitive
with free-knot splines when the temporal variation of $\phi$ is smooth
[\cite{rupp03}, Section~3.16]. \citet{wand00} reviews methods for
free-knot spline estimation; we do not consider this approach further.
To select $\tilde{\mathbf{t}}$ for a penalized spline, we consider a set of
values for $K$ with either equally-spaced or quantile-spaced $\tilde
{t}_k$'s, and select the partition that has low posterior mean deviance
and deviance information criterion (DIC) relative to the other
partitions considered [\cite{spie02}].

Following \citet{crai05}, without supplemental data, we complete the
Bayesian specification of (\ref{EmLRTP}) by placing vague priors on
$\beta_0$ and $\beta_1$, and a $\mathcal{N}(\mathbf{0},\sigma^2_{\bolds{\beta}}\bolds{\Omega}^{-})$ prior on $(\beta_2,\ldots,\beta_K)'$, where the
$(j,k)$th entry of $\bolds{\Omega}$ is defined as $|\tilde{t}_{j-1} -
\tilde{t}_{k-1}|^3$, for $j,k=2,\ldots,K$. In practice, we apply the
transformations $b_0 = \beta_0$, $b_1 = \beta_1$, $(b_2,\ldots,b_K)' =
\bolds{\Omega}^{1/2}(\beta_2,\ldots,\beta_K)'$, and $\sigma^{2}_{\mathbf{b}} =
\sigma^2_{\bolds{\beta}}$, and then complete the model specification by assuming
\begin{eqnarray}
\pi^*(b_k) & \equiv&
\mathcal{N}\bigl(b_k|0,10^4\bigr)\qquad \mbox{for } k=0,1,
\nonumber\\
\label{ELRTPPrior}
\pi^*(b_k|\sigma_{\mathbf{b}})& \equiv & \mathcal{N}
\bigl(b_k|0,\sigma _{\mathbf{b}}^{2}\bigr)\qquad \mbox{for }
k=2,\ldots,K \quad\mbox{and}
\\
\nonumber
\pi^*(\sigma_{\mathbf{b}}) & \equiv & \mathcal{U}(\sigma
_{\mathbf{b}}|0.01,100).
\end{eqnarray}
The prior in (\ref{ELRTPPrior}) smooths $\phi$ by preferring values
for $(b_2,\ldots,b_K)$ near zero {a priori}.

\subsection{GMC prior specification}\label{SSGMCP}
With supplemental data [$\mathbf{y}_0 = (y_{0,1},\ldots,\break y_{0,n_0})$, $\mathbf{t}_0  = (t_{0,1},\ldots,t_{0,n_0})$], we also assume
$y_{0,i_0}|t_{0,i_0} \sim \mathcal{N}(\phi_0(t_{0,i_0}),\sigma
^2_0)$, for $i_0 = 1,\ldots,n_0$. To ensure that $\phi$ and $\phi_0$
are analogous, we use the same partition $\tilde{\mathbf{t}}$ for $\phi_0$,
and model $\phi_0(t;\bolds{\beta}_0)$ using (\ref{EmLRTP}) by replacing
$\bolds{\beta}$ with $\bolds{\beta}_0$. We apply the transformation\vspace*{1.5pt} $b_{0,k}
= \beta_{0,k}$, for $k=0,1$, and $(b_{0,2},\ldots,b_{0,K})' = \bolds{\Omega}^{1/2}(\beta_{0,2},\ldots,\beta_{0,K})'$, and use the prior
specifications $\pi^*(b_{0,k}) \equiv\mathcal{N}(b_{0,k}|0,10^4)$, for
$k=0,1$, $\pi^*(b_{0,k}|\sigma_{\mathbf{b}_0}) \equiv\mathcal
{N}(b_{0,k}|0,\sigma_{\mathbf{b}_0}^{2})$ for $k=2,\ldots,K$, and $\pi
^*(\sigma_{\mathbf{b}_0}) \equiv\mathcal{U}(\sigma_{\mathbf{b}_0}|0.01,100)$.

To borrow flexibly for the shape of $\phi$ from the supplemental data
for $\phi_0$, we apply the GMC prior specification defined in (\ref
{EGenMixCommPr}) to $\mathbf{b}_{(-0)}=(b_1,\ldots,b_K)$ given $\mathbf{b}_{0,(-0)}=(b_{0,1},\ldots,b_{0,K})$. We assume
\begin{eqnarray*}
&&\pi(\mathbf{b}_{(-0)}|\mathbf{b}_{0,(-0)},\bolds{\iota}_{(-0)},\sigma_{\mathbf{b}}) \\
&&\qquad= \bigl[\pi^*(b_1)
\bigr]^{(1-\iota_1)} \bigl[\pi (b_1|b_{0,1})
\bigr]^{\iota_1} \prod_{k=2}^{K}
\bigl[\pi^*(b_k|\sigma_{\mathbf{b}}) \bigr]^{(1-\iota_k)} \bigl[
\pi(b_k|b_{0,k}) \bigr]^{\iota_k},
\end{eqnarray*}
where $\bolds{\iota}_{(-0)}$ = $(\iota_1,\ldots,\iota_K)$, and $\pi
^*(b_1)$, $\pi^*(b_k|\sigma_{\mathbf{b}})$ and $\pi^*(\sigma_{\mathbf{b}})$ are
defined in (\ref{ELRTPPrior}). We set $\pi(b_k|b_{0,k})$ $\equiv$
$\mathcal{N}(b_k |b_{0,k},\mathcal{R}_{\mathbf{b}}^{-1})$, for $k=1,\ldots
,K$, where $\mathcal{R}_{\mathbf{b}}$ is a prespecified \textit{large} value.
We then specify $\iota_k|\nu \stackrel{\mathrm{iid}}{\sim} \operatorname{Bern}(\nu)$ and
$\nu$ $\sim$ $\mathcal{B}(a_1,a_2)$ with fixed, known hyperparameters
$a_1$ and $a_2$ that reflect our prior opinion about the relevance of
the supplemental data for the shape of $\phi$.

To allow differential borrowing for the intercept versus the shape of
$\phi$, we specify an independent commensurate prior distribution for
$b_0$ given $b_{0,0}$ following (\ref{ESpSl2}). We define\vspace*{1pt}
$\pi(b_0|b_{0,0},\iota_0,\tau)\equiv[\pi
^*(b_0|b_{0,0},\tau)  ]^{(1-\iota_0)}[\pi
(b_0|b_{0,0})  ]^{\iota_0}$,
where $\pi^*(b_0|b_{0,0},\tau)\equiv\mathcal{N}(b_0|b_{0,0},\tau
^{-1})$ and $\pi(b_0|b_{0,0})\equiv\mathcal
{N}(b_0|b_{0,0},\mathcal{R}^{-1})$. We then specify $\tau \sim \mathcal{U}(s_l,s_u)$ and $\iota_0\sim\operatorname{Bern}(p_0)$,
where $s_l$,
$s_u$, $\mathcal{R}$ and $p_0$ are prespecified following the guidance
in Section~\ref{SSupp}. Taken together, the full posterior arises as
\begin{eqnarray}
&&p(\mathbf{b}, \mathbf{b}_0,
 \sigma_{\mathbf{b}}, \sigma_{\mathbf{b}_0}, \bolds{\iota}, \nu, \tau|
\mathbf{D}, \mathbf{D}_0) \nonumber\\
 &&\qquad\propto  L(\mathbf{b} | \mathbf{D}) L(
\mathbf{b}_0 | \mathbf{D}_0) \pi(b_0|b_{0,0},
\iota_0,\tau) \pi(\iota_0) \pi(\tau) \pi
^*(b_{0,0})
\nonumber
\\[-8pt]
\label{EGMCPost}
\\[-8pt]
\nonumber
&&\qquad\quad{}\times
 \pi(\mathbf{b}_{(-0)}|\mathbf{b}_{0,(-0)},\bolds{\iota
}_{(-0)},\sigma_{\mathbf{b}}) \pi(\bolds{\iota}_{(-0)}|\nu)
\pi(\nu) \pi ^*(\sigma_{\mathbf{b}})\nonumber \\
&&\qquad\quad{}\times\pi^*(\mathbf{b}_{0,(-0)} |
\sigma_{\mathbf{b}_0}) \pi ^*(\sigma_{\mathbf{b}_0}).\nonumber
\end{eqnarray}

\subsection{Simulation assessment}\label{SSSim}
\begin{figure}[b]

\includegraphics{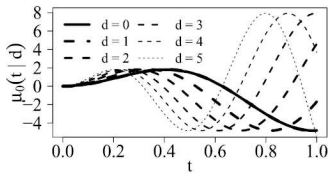}

\caption{Mean curves for the supplemental data for a subset of
discordance parameter ($d$) values. The mean curve for the primary data
is denoted by the solid line (i.e., $d=0$).}
\label{PSimFunc}
\end{figure}

We now investigate via simulation the borrowing properties of the GMC
prior model in (\ref{EGMCPost}) for nonparametric regression. To do
so, we sample $y_i|t_i \sim\mathcal{N} \lbrace\mu(t_i), \sigma
^{2}  \rbrace$, $i=1,\ldots,N$, where $\mu(t) = 5t\sin\lbrace5t
\rbrace$, and sample $y_{0,i_0}|t_{0,i_0},d \sim  \mathcal{N}\lbrace
\mu_0(t_{0,i_0}|d), \sigma_{0}^{2} \rbrace$, $i_0=1,\ldots,N_0$, where
$\mu_0(t|d) = (5+d)t\sin((5+d)t)$ and $d \in[0,5]$. Hence, the primary
data always have the same true mean structure, and the supplemental
data have a mean structure that deviates from that of the primary
according to the value of $d$, the discordance parameter. When $d = 0$,
the two curves are the same, and as $d$ increases, the supplemental
curve has increasingly greater curvature than the primary curve. Figure~\ref{PSimFunc} shows curves for selected values of $d$.

In each run, we sample $d$ uniformly from the set $\lbrace$0, 0.05,
0.10, 0.20, 0.35, 0.50, 0.75, 1, 1.50, 2, 3, 4, 5$\rbrace$. Given $d$,
we then generate primary and supplemental data sets with $N = N_0 =
50$, error variances $\sigma^{2} = \sigma_{0}^{2} = 1$, and equally
spaced $t$ and $t_0$ values over $[0,1]$. For each data set pair, $(\mathbf{D} = (\mathbf{y},\mathbf{t}),\mathbf{D}_0 =
(\mathbf{y}_0,\mathbf{t}_0))$, we fit the GMC
prior model developed in Section~\ref{SSGMCP}. We use the mLRTP model
defined in (\ref{EmLRTP}) with $K=10$ and $\tilde{\mathbf{t}}$ spaced
equally over $[0,1]$ for both $\phi_0(t_0;\bolds{\beta}_0)$ and $\phi(t;\bolds{\beta})$. After preliminary investigation, we set $s_l = 0$, $s_u =
2$, $\mathcal{R} = 2000$, $p_0 = 0.50$, and $\pi(\nu) \equiv\mathcal
{B}(\nu|0.50,0.50)$. The latter two choices represent an indifferent
prior opinion about the relevance of the supplemental data for both the
shape and the intercept of $\phi$. We choose to place an extremely
vague yet indifferent prior on $\nu$ to allow the primary and
supplemental data to have substantial influence on whether to borrow
for curve shape.

We used the {\tt R2jags} package to call {\tt JAGS} from {\tt R}, and
ran two chains for 20{,}000 iterations of burn-in, followed by 200{,}000
iterations for posterior estimation. We also fit the mLRTP model
defined in (\ref{EmLRTP}) with prior specifications described in (\ref
{ELRTPPrior}) to the primary data alone, as well as to the data set
obtained by simply pooling the primary and supplemental data. These
models feature lower MCMC autocorrelation, and thus required only two
chains with 2000 iterations of burn-in, followed by 20{,}000 iterations
for posterior estimation. These choices reflect preliminary
investigations to ensure acceptable MCMC convergence and relatively
small MC standard errors for the intercept and functional effect
coefficients. Estimation of the GMC model took about 45 seconds,
whereas estimation of each comparison model took about 5 seconds.

To evaluate the three models, we calculated four criteria at each run:
mean error (ME), root-mean-square error (RMSE), mean pointwise credible
interval width (CrIW), and mean pointwise coverage probability (CP).
Alternatively, we could calculate simultaneous confidence bands
following \citet{kriv10}, Section~3. We define the four criteria as
\begin{eqnarray*}
\operatorname{ME}\bigl(d^{(m)}\bigr) &=&  N^{-1}\sum
_{i=1}^{N} \bigl[\hat{\phi} \bigl(t_i|d^{(m)}
\bigr) - \mu \bigl(t_i|d^{(m)} \bigr) \bigr],
\\
\operatorname{RMSE}\bigl(d^{(m)}\bigr) &=& \Biggl\lbrace N^{-1}\sum
_{i=1}^{N} \bigl[ \hat{\phi} \bigl(t_i|d^{(m)}
\bigr) - \mu \bigl(t_i|d^{(m)} \bigr) \bigr]^2
\Biggr\rbrace^{{1}/{2}},
\\
\operatorname{CrIW}\bigl(d^{(m)}\bigr) &=& N^{-1}\sum
_{i=1}^{N} \bigl[\hat{\phi}_{0.975}
\bigl(t_i|d^{(m)} \bigr) - \hat{\phi}_{0.025}
\bigl(t_i|d^{(m)} \bigr) \bigr],
\\
\operatorname{CP}\bigl(d^{(m)}\bigr) &=& N^{-1}\sum
_{i=1}^{N} I \bigl\lbrace\mu \bigl(t_i|d^{(m)}
\bigr) \in \bigl[\hat{\phi}_{0.025} \bigl(t_i|d^{(m)}
\bigr), \hat{\phi}_{0.975} \bigl(t_i|d^{(m)}
\bigr) \bigr] \bigr\rbrace,
\end{eqnarray*}
for $m=1,\ldots,M$, where $\hat{\phi} (t|d^{(m)} )$ denotes
the posterior mean estimate for $\phi (t|d^{(m)} )$, and
$\hat{\phi}_{q} (t|d^{(m)} )$ denotes the $q$th quantile
posterior estimate of $\phi (t|d^{(m)} )$ in the $m$th run.
We then compare the sampling average of each criterion over the $M$
simulated data set pairs as a function of $d$, the discordance parameter.
Specifically, we visually compare Loess estimates of each evaluation
criteria as a function of the discordance parameter.
One simulation iteration takes about 60 seconds, which entails
generating a pair of data sets, fitting the three models, and then
calculating and saving the evaluation criteria, along with the sampled
value of $d$ in that run. We reduced overall computation time using the
{\tt snowfall} package for {\tt R} to conduct the simulation runs in
parallel. A {\tt R} program to implement this simulation is available
through the third author's software page \surl{http://www.biostat.umn.edu/\textasciitilde brad/software.\\html}.

\begin{figure}

\includegraphics{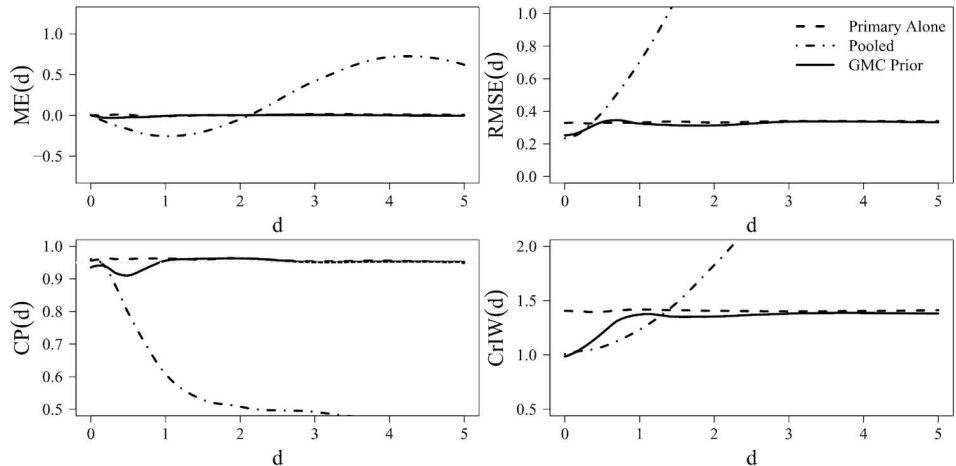}

\caption{Simulation assessment of the GMC prior spline model (solid)
versus a conventional spline model fit to the primary data alone
(dashed) and to the data set obtained by pooling the supplemental and
primary data (dot-dashed). These methods are compared on sampling
averages of mean error (top left), root-mean-square error (top right),
mean credible interval width (bottom right) and mean pointwise coverage
probability (bottom left) as a function of $d$, the discordance
parameter. All results are based on $M$ = 2000 runs.}\label{PSimRes}
\end{figure}

The results of our simulation investigation are illustrated in Figure~\ref{PSimRes}. Each panel shows the Loess estimates of the sampling
average of the corresponding evaluation criteria as a function of $d$
for three models: the GMC prior model (solid), the conventional model
fit to the primary data alone (dashed), and the conventional model fit
to the pooled data (dot-dashed). As expected, inferential properties
for the GMC prior model reflect those for the pooled approach under
true concordance (i.e., $d=0$) and approach primary data alone for
increasing degrees of discordance. The primary-alone approach has the
largest average CrIW and RMSE under true concordance, but its
properties do not deteriorate for increasing degrees of discordance
(because it does not acknowledge the supplemental data). By contrast,
the pooling approach has the smallest CrIW and RMSE under true
concordance, but its performance on all four evaluation criteria
deteriorates substantially as discordance increases.

The ME plot (top left panel) in Figure~\ref{PSimRes} demonstrates that
the GMC prior model has bias properties that interpolate the two
conventional approaches for near concordance, but near $d=1$ it learns
to effectively ignore the supplemental information. Furthermore, the
RMSE (top right panel) shows that the GMC prior model learns to borrow
strongly from the supplemental information when the sources are
commensurate (i.e., slight discordance), achieving RMSE similar to that
of the pooling approach. Similarly, the CrIW plot (bottom right panel)
shows for slight discordance that the credible intervals from the GMC
prior model are nearly as tight as those for the pooling approach.
Conversely, when the supplemental curve has greater oscillation than
the primary curve, the GMC prior approach has credible interval width
similar to those of the primary-alone approach. Last, the CP plot
(bottom left panel) shows that the mean pointwise coverage
probabilities for the GMC prior model are similar to those of the
conventional spline model fit to the primary data alone.

\subsection{Application: Liver imaging study}\label{SSRadOnc}
We now apply our GMC prior model structure to estimate the temporal
features of each perfusion characteristic in cancerous liver tissue.
Recall that similar information derived from noncancerous tissue is
potentially valuable supplemental information because the shape of the
average CTp curve in noncancerous regions may provide relevant
information about the shape of the corresponding average CTp curve in
cancerous regions. The data consist of 7 to 13 readings acquired at
times between 11 and 100 seconds after contrast injection in 0 to 2
cancerous and noncancerous regions of interest (ROIs) among 16
individuals. For each perfusion characteristic, there are 687 total
readings from fifteen individuals who each contribute $N_i =
11\mbox{--}13$
readings in $M_{T,i} = 1\mbox{--}2$ cancerous and $M_{N,i} = 1\mbox{--}2$ noncancerous
ROIs, and one individual who contributes 7 readings in 1 noncancerous ROI.

Individual acquisition times $t_{i,\ell}$, $\ell=1,\ldots,N_i$, are
necessarily identical for all ROIs and all perfusion characteristics.
Thus, we let $y_{r,i,j,\ell}$, for $j$ = $1,\ldots,M_{r,i}$, denote a
reading at time $t_{i,\ell}$. We denote the average CTp curve for a
given perfusion characteristic in cancerous tissue and noncancerous
tissue by $\phi_{T}(t)$ and $\phi_{N}(t)$, respectively. We assume each
individual's CTp curve deviates from the average CTp curve smoothly
over time, and we denote these deviation curves by $\psi_{r,i}(t)$,
$r= T$, $N$ and $i = 1,\ldots,16$. We then model
\begin{equation}
\label{ECTpModel}
y_{r,i,j,\ell} \sim\mathcal{N} \bigl\lbrace\phi_{r}(t_{i,\ell})
+ \psi _{r,i}(t_{i,\ell}), \sigma^2_{e,r}
\bigr\rbrace.
\end{equation}
We use the mLRTP model defined in (\ref{EmLRTP}) for $\phi_{r}$ and
$\psi_{r,i}$, parametrized by $\bolds{\beta}_{r}$ and $\bolds{\alpha}_{r,i}$,
respectively. Therefore, the first derivative of the average CTp curve is
\begin{equation}
\label{ELRTPDeriv}
\phi'_{r}(t;\bolds{\beta}_r)
= \beta_{r,1} + \sum_{k=2}^{K}
\operatorname{sign}(t-\tilde{t}_{k-1}) 3\beta_{r,k}(t-\tilde{t}_{k-1})^2,\qquad
r=N,T.
\end{equation}
To shrink the individual deviation curves $\psi_{r,i}(t;\bolds{\alpha}_{r,i})$ toward the average CTp curve $\phi_r(t;\bolds{\beta}_r)$, we
assume $\alpha_{r,i,k} \sim\mathcal{N}(0,\sigma^2_{\mathbf{a},r,i})$, for
$k=0,\ldots,K$, $r=N,T$ and $i=1,\ldots,16$.

For our\vspace*{1pt} conventional model, we use the prior specifications on $\bolds{\beta}_{N}$ and $\bolds{\beta}_{T}$ described in (\ref{ELRTPPrior}),
and specify vague $\mathcal{N}(0,10^4)$ priors for $b_{N,0}$ and
$b_{T,0}$. For the GMC prior model, we apply the prior developed in
Section~\ref{SSGMCP} for $\bolds{\beta}_{T}$ given $\bolds{\beta}_{N}$. That
is, we specify a GMC prior for $\bolds{\beta}_{T,(-0)} = (\bolds{\beta
}_{T,1},\ldots,\bolds{\beta}_{T,K})$ given $\bolds{\beta}_{N,(-0)} = (\bolds
{\beta}_{N,1},\ldots,\bolds{\beta}_{N,K})$ and an independent commensurate
prior for $b_{T,0}$ given $b_{N,0}$. This choice enables differential
borrowing from the noncancerous tissue data for the intercept versus
the curve shape parameters, and simultaneously enables borrowing
similar amounts among curve shape parameters. Thus, if the average CTp
curve in noncancerous regions differs from the average CTp curve in
cancerous regions only for the intercept, the model still permits
borrowing for the CTp curve shape. Last, we place vague $\mathcal
{U}(0.01,100)$ priors on each of the standard deviation parameters
($\sigma_{e,r}$, $\sigma_{\mathbf{b},r}$, $\sigma_{\mathbf{a},r,i}$).

We let $t$ $\in$ $(0,1]$ by taking $t = \frac{t^*-t^*_{\min}}{t^*_{\max}
- t^*_{\min}}$, where $t^*$ is the original timescale. Preliminary
analysis of the noncancerous regions was used to specify the
hyperparameters $\mathcal{R} = 500$, $s_l = 0.01$, $s_u = 0.50$, $p_0 =
0.10$, $a_1 = 0.10$ and $a_2 = 0.90$. To select a partition, we fit our
conventional model (i.e., independent spline models for the average CTp
curve in each tissue type) for each perfusion parameter using $K$ = 5,
10, 15 and 25 with $\tilde{\mathbf{t}}$ equally spaced over $[0,1]$, and
using $K$ = 5, 10 and 15 with $\tilde{\mathbf{t}}$ placed at equally spaced
quantiles of the observed acquisition times. The model using $K=10$
with quantile-spaced knots resulted in low deviance and low DIC
relative to models with the other partitions for all three perfusion
characteristics, so we chose to conduct our analysis using this
partition. We estimated the conventional model using 40{,}000 posterior
samples after 2000 burn-in samples from two MCMC chains. We estimated
the GMC model posterior using 200{,}000 posterior samples after 20{,}000
burn-in samples from two MCMC chains; there was greater autocorrelation
for the basis coefficient parameters than the conventional model.

\begin{table}[b]
\tabcolsep=0pt
\caption{Posterior borrowing parameter estimates (i.e., $\iota_k$'s).
$\iota_0$ corresponds to the CTp curve intercept, and the remaining
$\iota_k$'s correspond to CTp curve shape. Values near one indicate
strong borrowing}\label{TCTpResults}
\begin{tabular*}{\tablewidth}{@{\extracolsep{\fill}}lccccccccccc@{}}
\hline
\textbf{Perfusion characteristic} & $\bolds{\iota_0}$ & $\bolds{\iota_1}$ &
$\bolds{\iota_2}$ & $\bolds{\iota
_3}$ & $\bolds{\iota_4}$ & $\bolds{\iota_5}$ & $\bolds{\iota_6}$ &
$\bolds{\iota_7}$ & $\bolds{\iota_8}$ &
$\bolds{\iota_9}$ & $\bolds{\iota_{10}}$ \\
\hline
PS & 0.00 & 0.23 & 0.00 & 0.00 & 0.24 & 0.43 & 0.24 & 0.05 & 0.15 &
0.03 & 0.34 \\
BV & 0.20 & 0.92 & 0.95 & 0.92 & 0.82 & 0.86 & 0.90 & 0.82 & 0.88 &
0.96 & 0.96 \\
BF & 0.02 & 0.96 & 0.98 & 0.92 & 0.92 & 0.93 & 0.94 & 0.92 & 0.96 &
0.98 & 0.99 \\
\hline
\end{tabular*}
\end{table}

Table~\ref{TCTpResults} reports the posterior mean borrowing
parameters (i.e., $\iota_k$'s) from the GMC model for the three
perfusion characteristics. Values near 1 indicate strong borrowing for
the corresponding basis parameter. The first column reports the
posterior mean for $\iota_0$, which corresponds to the intercept. PS
and BF show virtually zero borrowing for the intercept, whereas BV
shows little borrowing. The remaining columns illustrate borrowing for
the basis parameters that control the shape (and thus, first
derivative) of the CTp curve, which we have restricted to have similar
magnitude by assuming $\iota_k|\nu  \sim  \operatorname{Bern}(\nu)$, for $k =
1,\ldots,10$, {a priori}. For PS, these parameters all have
posterior means less than 0.43, and many less than 0.25, which
indicates little borrowing for CTp curve shape. In contrast, these
parameters all exceed 0.82 for BV, and 0.92 for BF. Thus, for BF and BV
the GMC model borrows substantially from the supplemental information
in noncancerous tissue for CTp curve shape in cancerous tissue.

\begin{figure}

\includegraphics{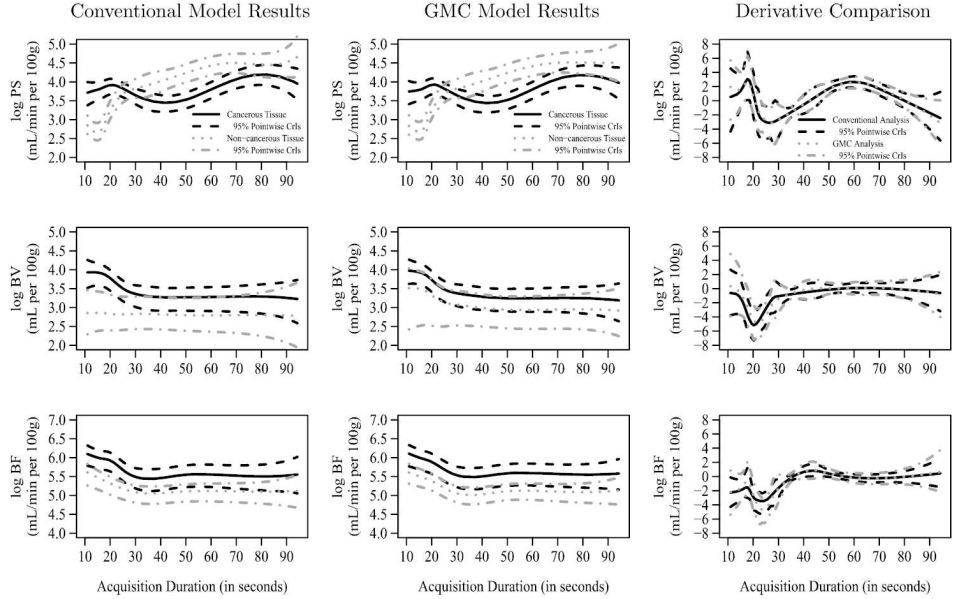}

\caption{The first column displays posterior mean CTp curves from the
conventional model analysis for cancerous (solid black) and
noncancerous regions (dotted grey), along with 95\% pointwise CrIs.
The second column displays the same results for the GMC model analysis.
The third column compares the posterior mean first derivative of the
CTp curve in cancerous tissue from the conventional model (solid black)
and the GMC prior model (dotted grey), along with 95\% pointwise
CrIs.}\label{PCTpResults}
\end{figure}

The first row in Figure~\ref{PCTpResults} indicates that both the
shape and intercept of the PS curve differ substantially by tissue type
and that the results of the conventional model analysis (first column)
and the GMC prior model analysis (second column) are virtually
indistinguishable. Consequently, the first derivative of the CTp curve
in cancerous tissue (third column) is estimated with similar precision
in either the conventional or GMC model analysis. In fact, the GMC
prior model results in 95\% pointwise credible intervals (CrIs) that
are 4\% wider on average than those of the conventional model analysis.
Regardless, the PS CTp curve in cancerous tissue has not stabilized
after 95 seconds, so longer acquisition durations for PS appear
necessary, which is consistent with the findings by \citet{ng13}.

The second row in Figure~\ref{PCTpResults} corresponds to BV. Here the
GMC prior model, relative to the conventional model, results in a
noncancerous CTp curve that is more similar for both the intercept and
the shape to that of the cancerous tissue CTp curve. Moreover, the GMC
model gains precision over the conventional model for the first
derivative estimate of the cancerous tissue CTp curve via borrowing
from the noncancerous tissue CTp curve shape. The average 95\%
pointwise CrIs for the GMC model are 18\% tighter than those of the
conventional model. The third row corresponds to BF and illustrates
that the posterior CTp curve estimates are similar for the two modeling
approaches. However, the first derivative of the cancerous tissue CTp
curve is estimated more precisely for the GMC model than the
conventional model, resulting in 27\% tighter 95\% pointwise CrIs on
average. For both BF and BV, we can infer a shorter acquisition
duration using the GMC prior model than using the conventional model,
although we did not impose precise stability criteria here.

\section{GMC priors in semiparametric survival analysis}\label{SHazReg}
We now describe a flexible semiparametric survival model and then apply
our GMC prior technology to the colorectal cancer clinical trials data
described in Section~\ref{SIntro}. For simplicity, assume each
observation consists of a possibly right-censored time $t \in(0,1]$, a~binary event indicator $c$ and a binary treatment indicator~$z$.

\subsection{Piecewise-exponential proportional hazards model}\label{SSPwExp}
A flexible survival model commonly favored by Bayesians is the
piecewise-exponential proportional hazards model, which is constructed
by partitioning the time axis into $K$ intervals $\tilde{\mathbf{t}} =
(0=\tilde{t}_0<\tilde{t}_1<\cdots<\tilde{t}_{K-1}<\tilde{t}_K=1)$ and
assuming the baseline hazard is constant in each interval [cf. \cite
{ibra01}, Section~3.1]. Under this model, the likelihood for an
observation $(t,c,z)$ is given by $h(t|z;\bolds{\gamma},\rho)^{c}\exp
\lbrace- \int_{0}^{t} h(s|z;\bolds{\gamma},\rho)\,ds  \rbrace$, where
\begin{equation}
\label{EPWE}
h(t|z;\bolds{\gamma},\rho) = \exp(\gamma_k + \rho z)
\qquad\mbox{for } t \in I_k = (\tilde{t}_{k-1},
\tilde{t}_k ], k=1,\ldots,K.
\end{equation}
Thus, $h(t|z;\bolds{\gamma},\rho)$ is assumed to be piecewise-constant,
where $\gamma_k$ denotes the log-hazard in the $k$th interval of the
time-axis partition $\tilde{\mathbf{t}}$ for treatment assignment $z=0$,
and $\rho$ denotes the log-hazard ratio for treatment assignment $z=1$
relative to $z=0$. Following \citet{ibra01}, we select $K$ using a DIC
comparison over a small set of partitions. To resist overfitting, we
specify a correlated prior process for $\pi^*(\bolds{\gamma})$. We focus
on the random-walk prior process for $\pi^*(\bolds\gamma)$ introduced by
\citet{fahr01}. Formally, we specify
\begin{eqnarray}
\pi^*(\gamma_1) &\equiv &
\mathcal{N}\bigl(\gamma_1|0,10^4\bigr),
\nonumber\\
\label{ERWPP}
\pi^*(\gamma
_k|\gamma_{k-1},\sigma_{\bolds{\gamma}}) & \equiv & \mathcal{N}
\bigl(\gamma_k|\gamma _{k-1},\sigma_{\bolds{\gamma}}^{2}
\bigr)\qquad \mbox{for } k=2,\ldots,K \quad\mbox{and}
\\
\nonumber
 \pi^*(\sigma_{\bolds{\gamma}})  &\equiv & \mathcal {U}(
\sigma_{\bolds{\gamma}}|0.01,100).
\end{eqnarray}
The prior process in (\ref{ERWPP}) smooths adjacent $\gamma_k$'s
toward each other by assuming their first differences are exchangeable.
We also assume $\rho$ is independent of $\bolds{\gamma}$ {a~priori},
and specify a vague $\mathcal{N}(0,10^4)$ prior. The model defined in
(\ref{EPWE}) is useful in the absence of supplemental data, and the
parameter space can be decomposed into separable subsets, $\bolds{\gamma}$
characterizes the baseline hazard and $\rho$ characterizes the
treatment effect.

\subsection{Application: Colorectal cancer trials}\label{SSColCan}
Extending \citet{hobb13}, we apply our GMC prior model structure to
supplement the inference on the progression-free survival (PFS) curve
among all three regimens in the trial reported by \citet{gold04} using
the historical information on the IFL regimen from the trial reported
by \citet{salt00}. We use the piecewise-exponential
proportional-hazards model to estimate the PFS curves for the two
trials from these data. During the first two years of follow-up, the
primary (Goldberg) data set contains 197 progression events among 211
persons treated with IFL, 190 events among 216 persons treated with
FOLFOX, and 189 events among 206 persons treated with IROX. For the
primary data, we have two binary treatment indicators for assignment to
the FOLFOX ($z_{F}$) and IROX ($z_{I}$) regimens, respectively, so
following (\ref{EPWE}) we model
\[
h(t|\mathbf{z};\bolds{\gamma},\bolds{\rho}) = \exp(\gamma_k +
\rho_{F} z_{F} + \rho_{I} z_{I})
\qquad\mbox{for } t \in(\tilde{t}_{k-1},\tilde{t}_{k}],
\]
where $k=1,\ldots,K$. The supplemental (Saltz) data contain 172
progression events during the first two years of follow-up among 224
persons treated with IFL. The model for the supplemental data is
defined analogously, though $z_{F} = z_{I} = 0$ for all the
supplemental observations. Therefore, the supplemental hazard model is
completely parametrized by $\bolds{\gamma}_0$, which, as required, is
analogous to $\bolds{\gamma}$.

We specify a GMC prior to flexibly borrow strength from the
supplemental information on the IFL regimen. Namely, we specify the
random-walk prior process defined by (\ref{ERWPP}) for $\pi^*(\bolds{\gamma}_0)$, and apply the GMC prior structure on $\bolds{\gamma}$. In
this\vspace*{1pt} setting, the prior specification follows as
$\gamma_1|\gamma_{0,1},\iota_1 \sim[\mathcal{N}(\gamma
_1|0,10^4) ]^{1-\iota_1}  [\mathcal{N}(\gamma
_1|\gamma_{0,1},\mathcal{R}_{\bolds{\gamma}}^{-1}) ]^{\iota
_1}$, $\gamma_k|\gamma_{k-1}, \gamma_{0,k},\iota_k$, $\sigma_{\bolds{\gamma}}
\sim[\mathcal{N}(\gamma_k| \gamma_{k-1},\break\sigma^{2}_{\bolds{\gamma}}) ]^{1-\iota_k} [\mathcal{N}(\gamma
_k| \gamma_{0,k},\mathcal{R}_{\bolds{\gamma}}^{-1}) ]^{\iota
_k}$, for $k=2,\ldots,K$, and $\iota_k|\nu_{\bolds{\gamma}} \sim
\operatorname{Bern}(\nu_{\bolds{\gamma}})$, for $k=1,\ldots,K$,
where $\mathcal{R}_{\bolds{\gamma}}$ is prespecified. Next, we place a
$\mathcal{B}(a_1,a_2)$ prior on $\nu_{\bolds{\gamma}}$, where $a_1$ and
$a_2$ are also prespecified. Last, we place vague $\mathcal
{U}(0.01,100)$ priors on each of the standard deviation parameters
(i.e., $\sigma_{\bolds{\gamma}}$ and $\sigma_{\bolds{\gamma}_0}$) and vague
$\mathcal{N}(0,10^4)$ priors on the treatment effect parameters (i.e.,
$\rho_{F}$ and $\rho_{I}$).

We transformed the timescale so that $t$ $\in$ $(0,1]$, then selected
$\mathcal{R}_{\bolds{\gamma}} = 10{,}000$ and specified a vague $\mathcal
{B}(0.10,0.90)$ prior for $\nu_{\bolds{\gamma}}$, which represents a
vague, yet skeptical prior opinion about the relevance of the
supplemental data. For estimation, we used 200{,}000 posterior samples
for estimation after 50{,}000 iterations of burn-in from two MCMC chains.
For comparison, we also fit conventional piecewise-exponential
proportional hazards models with the same time-axis partition and
random-walk prior process to the primary (Goldberg) data alone,
supplemental (Saltz) data alone, and the data set obtained by naively
pooling the two sources of information. To estimate these conventional
models, we used 20{,}000 posterior estimation draws after 2000 burn-in
draws from two MCMC chains, again the shorter chain length justified by
faster convergence and lower post-convergence autocorrelations.

\begin{figure}[b]

\includegraphics{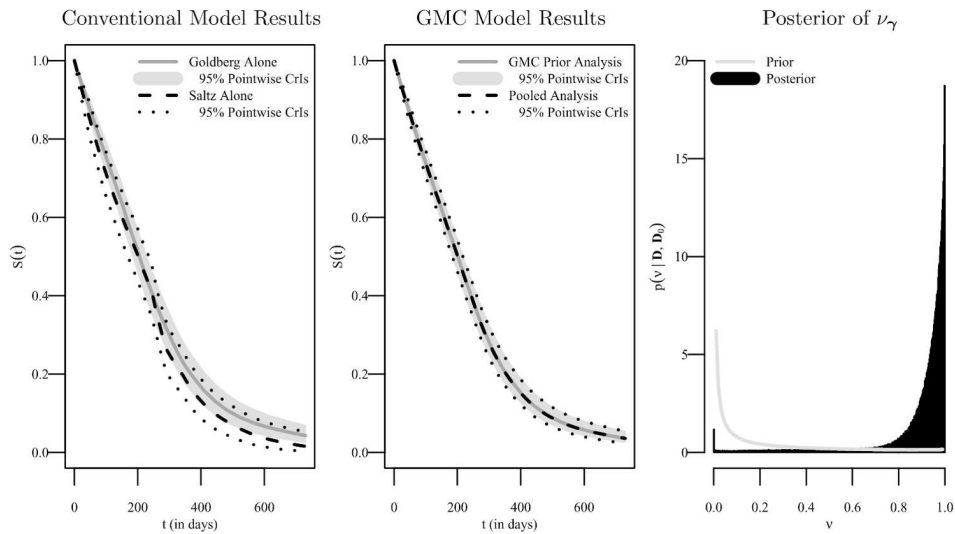}

\caption{The left panel displays the posterior mean PFS curve for the
IFL regimen along with 95\% pointwise CrIs from the conventional
piecewise-constant-hazard analysis of the primary (Goldberg) data alone
and the supplemental (Saltz) data alone. The middle panel displays the
same results from the GMC prior model analysis and the conventional
model analysis of the dataset obtained by pooling the two sources of
information. The third panel displays a posterior histogram for $\nu
_{\bolds{\gamma}}$, which controls borrowing for the baseline hazard among
trials. For reference, the prior assigned to $\nu_{\bolds{\gamma}}$ is
also displayed (grey line).}\label{PColCanPlots}
\end{figure}

We begin with a comparison of the estimated PFS curves for the IFL
regimen from the four analyses. The left panel in Figure~\ref{PColCanPlots} indicates that the estimated PFS curve from the
Goldberg data (solid, with shaded grey 95\% pointwise CrIs) overlaps
substantially with the PFS curve estimated from the Saltz data (dashed,
with dotted 95\% pointwise CrIs), suggesting the Saltz data provide
relevant information. The middle panel in Figure~\ref{PColCanPlots}
illustrates that the GMC analysis (solid, with shaded grey 95\%
pointwise CrIs) results in a PFS estimate nearly indistinguishable from
that of the conventional analysis of the pooled data set (dashed, with
dotted 95\% pointwise CrIs), though shifted very slightly toward the
primary information. The GMC prior model achieved a PFS curve estimate
with noticeably greater precision than that of the conventional
analysis that ignores the supplemental data without requiring bold {a priori} assumptions regarding the relevance of the Saltz data.
Finally, the rightmost panel in Figure~\ref{PColCanPlots} displays a
posterior histogram of $\nu_{\bolds{\gamma}}$ that has much of its mass
shifted toward one. In fact, the $\iota_k$'s had posterior means of at
least 0.87, with an average of 0.94, indicating that the GMC prior
model has learned that the supplemental FOLFOX data from the Saltz
trial are indeed relevant.

\begin{table}
\caption{Hazard-ratio estimates for the FOLFOX and IROX regimens
relative to the IFL regimen, and estimates of median days to disease
progression for each treatment regimen from the four analyses of the
Goldberg data and Saltz data}
\label{THRandMSTab}
\begin{tabular*}{\tablewidth}{@{\extracolsep{\fill}}lccc@{}}
\hline
\multicolumn{1}{@{}l}{\textbf{Drug regimen}} & \textbf{IFL} & \textbf{FOLFOX} & \textbf{IROX} \\
\hline
\multicolumn{1}{@{}l}{\textit{Analysis}} & &\multicolumn{2}{c@{}}{\textit{Hazard
ratios (\textup{95}\% CrIs)}} \\
Goldberg alone & -- & 0.70 (0.57, 0.86) & 0.92 (0.75, 1.12) \\
Pooled analysis & -- & 0.66 (0.56, 0.79) & 0.87 (0.73, 1.04) \\
GMC prior analysis & -- & 0.66 (0.56, 0.79) & 0.87 (0.73, 1.04) \\[6pt]
& \multicolumn{3}{c@{}}{\textit{Median days to disease progression (\textup{95}\% CrIs)}} \\
Saltz alone & 200 (168, 236) & -- & -- \\
Goldberg alone & 205 (182, 227) & 262 (235, 293)& 217 (194, 242)\\
Pooled analysis & 200 (182, 217) & 263 (245, 286) & 221 (201, 240)\\
GMC prior analysis & 200 (184, 216) & 265 (245, 286) & 221 (203, 240) \\
\hline
\end{tabular*}
\end{table}

Turning to the comparative effectiveness of three regimens, Table~\ref{THRandMSTab} shows that the hazard ratios (95\% CrIs) from the GMC
prior model analysis for FOLFOX and IROX versus IFL are 0.66
(0.56,0.79) and 0.87 (0.73,1.04), respectively. The
hazard ratios from the conventional analysis of the
Goldberg data alone are slightly larger with notably wider CrI widths,
0.70 (0.57,0.86) for FOLFOX and 0.92 (0.75,1.12) for IROX. By contrast,
the hazard-ratio estimates and CrI widths from the conventional
analysis of the pooled data set are indistinguishable from the GMC
prior model analysis. Table~\ref{THRandMSTab} also contains the
posterior estimates from each analysis of median PFS for the IFL,
FOLFOX and IROX regimens. The conventional analysis of the Saltz data
alone estimates median PFS in the IFL regimen to be 200 (168, 236)
days, broadly similar to the estimate of 205 (182, 227) days provided
by the conventional analysis of the Goldberg data alone. The estimates
of median days to disease progression for each regimen from the pooled
analysis of these data are also nearly identical to the estimates from
the GMC prior analysis, namely, 200, 265 and 221 days for the IFL,
FOLFOX and IROX regimens, respectively. The estimates from the GMC
prior model are consistent with the results reported by \citet{gold04},
who found FOLFOX to be the superior regimen, and significantly better
than IFL. In addition, the GMC prior analysis yields stronger evidence
that IROX may be better than IFL for PFS, though the difference remains
statistically insignificant.

\section{Discussion and future work}\label{SDisc}
Our proposed methods for prior specification in functional and survival
data models with penalized splines facilitate data-dependent borrowing
that is robust to biased estimation of primary effects when conflict
among information sources occurs. The simulation study illustrates the
beneficial flexible borrowing properties the proposed methods offer.
The application in perfusion CT illustrates potential gains in CTp
curve estimation from using supplemental data collected concurrently.
By contrast, the colorectal cancer application illustrates the use of
these methods for semiparametric survival modeling to flexibly borrow
from supplemental data on a control therapy collected in a previous
trial. A hierarchical model with the proposed structure allows the
degree of borrowing to be estimated differentially for each feature
(e.g., CTp curve intercept versus shape). The amount of strength being
borrowed between sources reflects the evidence of commensurability for
that feature. When substantial evidence indicates that sources differ
for a feature, the proposed method will learn to effectively ignore the
supplemental data for that feature, yet possibly still borrow strength
for another feature. For minor discordance or concordance, the proposed
method also facilitates partial (rather than full) pooling of
information from the supplemental source, hence a modest yet
justifiable gain in efficiency. Our general modeling strategy enables
data-driven estimation of heterogeneity among information sources for
complex functional relationships, and thus provides a sensible and
justifiable synthesis of clinical information.

In future work, we plan to extend the commensurate prior approach to
settings that have \textit{multiple} supplemental sources of information.
This setting substantially complicates the construction of a
hierarchical model that facilitates flexible borrowing for each
supplemental source. Ideally, the method will facilitate data-dependent
differential borrowing from various sources, learning from which to
borrow and which to ignore as primary data accumulate. We also hope to
develop concise, interpretable summaries that quantify the amount of
strength being borrowed from each information source, thereby allowing
the use of these models in an adaptive trial [see \cite{hobb13}]. We
will also consider using posterior summaries of $\nu$ in (\ref{EGenMixCommPr}) to determine whether the curves are commensurate
among information sources, for example, by assessing $\operatorname{Pr}(\nu> 0.80|\mathbf
{D},\mathbf{D}_0) > 0.90$. Furthermore, we are currently studying the
properties of the piecewise-exponential model used in the colorectal
cancer application, which relies on a simple, yet flexible
piecewise-constant assumption for the baseline hazard. We are
developing extensions that use a piecewise-linear model for the
baseline hazard function, relax the proportional hazards assumption,
and allow functional covariate effects with shape constraints (e.g.,
monotonicity).

\section*{Acknowledgments}
We thank the Associate Editor and referees for their well thought-out
and detailed comments during revision that substantially improved the
clarity of presentation in this manuscript and the model we use in the
CTp analysis.

%





\printaddresses
\end{document}